\documentclass[prl,onecolumn,showpacs,preprintnumbers,amsmath,amssymb,floatfix]{
revtex4}

\usepackage[dvips]{graphics,graphicx,color}

\begin{document}

\title{First-Principles Study of a Positron Immersed in an Electron
Gas}

\author{N.\ D.\ Drummond}

\author{P.\ L\'opez\ R\'{\i}os}

\author{C.\ J.\ Pickard} \altaffiliation[Current address: ]{University College
London, Gower Street, London WC1E 6BT, United Kingdom.}

\author{R.\ J.\ Needs}

\affiliation{TCM Group, Cavendish Laboratory, University of Cambridge, J.\ J.\
Thomson Avenue, Cambridge CB3 0HE, United Kingdom}

\date{\today}

\begin{abstract} 
  Calculations of the relaxation energy, contact pair-correlation
  function, and annihilating-pair momentum density for a single
  positron immersed in a homogeneous electron gas are presented.
  We achieve an accurate description of the electron-positron
  correlation effects by working in the reference frame in which the
  positron is stationary and using a mean-field approach based on
  single-component density functional theory.  Our positron
  relaxation energies and annihilation rates are similar to those from
  the best existing many-body calculations.  Our annihilating-pair
  momentum densities are significantly different from previous data,
  and include a ``tail'' beyond the Fermi edge.
\end{abstract}


\maketitle

Quantum impurity problems are of widespread interest in condensed
matter physics.  Here we study a positron in a homogeneous electron
gas (HEG) as an example of an impurity problem in which a particle of
one species is immersed in a fluid of another species, and where
quantum effects are important for both the host and impurity.
Examples of other problems of this type are the Mahan exciton in
semiconductor physics, in which a hole in the valence band interacts
with an electron gas in the conduction band (or an electron interacts
with a hole gas), and an impurity atom in a Bose-Einstein condensate.
Our approach requires knowledge of the explicit interaction between
the host and impurity particles and the availability of a reasonably
accurate mean-field description of the host, such as the Kohn-Sham
density functional theory (DFT) \cite{hohenberg_1964,kohn_1965} of
electrons or the Gross-Pitaevskii equation
\cite{gross_1961,pitaevskii_1961} for Bose-Einstein condensates.  An
important characteristic of our approach is the inclusion of
nonadiabatic impurity-host effects.  We have chosen to present results
for the positron problem, firstly, because they can be compared with
data in the literature to gauge their accuracy and, secondly, because
nonadiabatic effects are expected to be important as the electron and
positron are of equal mass.  Applications of our method to other
problems such as those mentioned above require only straightforward
modifications.

On entering condensed matter, a low-energy positron thermalizes
rapidly and may be trapped by open-volume defects such as vacancies,
where the nuclear repulsion is weak.  The positron lifetime is
measured as the time interval between the detection of a photon
emitted in the $\beta^+$ radioactive decay that produces the positron
and the detection of the two 0.511 MeV photons emitted when the
positron annihilates an opposite-spin electron
\cite{krause_rehberg}. The lifetime is characteristic of the defect at
which the positron settles, and positron annihilation spectroscopy is
an important, nondestructive technique for characterizing open-volume
defects.  Measuring the Doppler broadening of the annihilation
radiation or the angular correlation between the two 0.511 MeV photons
yields information about the momentum density of the electrons in the
presence of the positron.  Modeling is then required to extract
information about the unperturbed system from the experimental data.

The most widely used method for modeling positrons in real materials
is two-component DFT \cite{boronski_1986}, in which the interaction
between the electrons and the positron is described by a functional of
the densities of the electron and positron components.  Within the
local density approximation (LDA), this functional is obtained
directly from the relaxation energy $\Delta \Omega$ of a positron in a
HEG, which is the difference between the energy of a HEG with and
without a positron immersed in it.  Equivalently, $\Delta \Omega$ is
the electron-positron correlation energy.

Two-component DFT can lead to accurate electron and positron
densities, but the DFT orbitals yield very poor annihilating-pair
momentum densities (MDs) $\rho(p)$ and electron-positron
pair-correlation functions (PCFs) $g(r)$ because they do not include
the full effects of the strong electron-positron correlation
\cite{boronski_1986,puska_1994}.  The results can, however, be
substantially improved by correcting the calculated PCFs and MDs using
accurate data for a positron in a HEG \cite{boronski_1986,puska_1994}.
The effective electron density felt by the positron, and hence the
annihilation rate, is proportional to the contact PCF between the
positron and the electrons, $g(0)$.  The annihilation rate for
a positron immersed in a paramagnetic HEG is \cite{rs_and_au} $\lambda
= 3 g(0)/(4c^3r_s^3)$, where $r_s$ is the electron density parameter,
and $c$ is the speed of light in vacuo. If the electron and positron
motions were uncorrelated then $g(0)$ would be unity, but the strong
Coulomb attraction leads to much larger values, particularly at low
densities, where a bound state may be formed.  Together, our results
for $\Delta \Omega$, $g(0)$, and $\rho(p)$ permit the construction of
a two-component DFT within the LDA for a positron in a HEG, which in
turn will enable the calculation of the annihilation rates and MDs
used to interpret the results of positron annihilation experiments.

We follow the suggestion of Leung \textit{et al.}\ \cite{leung} that
it is useful to describe a positron in a HEG using the set of electron
positions relative to the position of the positron, $\{ {\bf x}_i\}$.
Neglecting the center-of-mass motion, which is zero in the ground
state, the Hamiltonian becomes \cite{rs_and_au}
\begin{equation} \hat{H} = -\sum_i \left[ \nabla_i^2 - v_{\rm I}({\bf
x}_i) \right] + \sum_{j>i} \left[ v({\bf x}_i-{\bf x}_j) - \nabla_i
\cdot \nabla_j \right],
\label{eqn:trans_hamilt}
\end{equation}  where $v_{\rm I}$ is the interaction between the
positron impurity and the electrons and $v$ is the interaction between
the electrons.  In this reference frame the Hamiltonian describes
interacting fermions of mass $1/2$ a.u.\ and charge $-1$ a.u.\ with a
fixed positive charge of magnitude 1 a.u.\ at the origin, and an extra
attractive interaction $\hat{V}_{\rm e} = -\sum_{j>i} \nabla_i \cdot
\nabla_j$, which resembles the mass-polarization term encountered in
atomic physics when transforming to the center-of-mass frame.  The
advantage of this formulation is that the mean-field
single-determinant approximation for the solution to Eq.\
(\ref{eqn:trans_hamilt}) includes explicit electron-positron
correlation via the $v_{\rm I}$ term, whereas in the laboratory frame
the ground state consists of a completely delocalized positron and a
HEG\@.  Leung \textit{et al.}\ \cite{leung} did not, however,
calculate the single-determinant approximation to the solution of Eq.\
(\ref{eqn:trans_hamilt}), but instead they neglected $\hat{V}_{\rm e}$
and noted that the resulting Hamiltonian is the same as for a fixed
``proton'' in an ``electron gas'' with particle mass $1/2$ a.u.
Unfortunately this approximation is a gross violation of the
properties of the system, as the magnitude of the expectation value of
the neglected term is approximately equal to the kinetic energy of the 
system in the laboratory frame.

We have formulated a mean-field theory for the problem in which the
ground-state charge density of the Hamiltonian of Eq.\
(\ref{eqn:trans_hamilt}) is approximated by that arising from a single
determinant of orbitals $\phi_i({\bf x})$, and the
exchange-correlation effects between the electrons are described by a
density functional $E_{\rm xc}[n]$.  The resulting mean-field
equations are
\begin{eqnarray} & \left[ -\nabla^2 + \int n({\bf x}^\prime)v({\bf
      x}-{\bf x}^\prime) \, d{\bf x}^\prime + v_{\rm I}({\bf x}) +
    \frac{\delta E_{\rm xc}[n]}{\delta n({\bf x})} \right] \phi_i({\bf
    x}) & \nonumber \\ & {} + \sum_j f_j \langle \phi_j | \nabla
  \phi_i \rangle \cdot \nabla \phi_j({\bf x}) = {\cal E}_i \phi_i({\bf
    x}), & \label{eqn:ks_eqns}
\end{eqnarray} where $n({\bf x})$ is the electron density, $f_i$ is
the occupation of orbital $i$ and ${\cal E}_i$ is the orbital
eigenvalue.  All our DFT calculations were performed with orbitals at
zero wave vector and with closed-shell configurations, in which case
(i) the center-of-mass kinetic energy is zero in the ground state and
(ii) the direct term in the expectation value of the extra interaction
$\hat{V}_{\rm e}$ is zero, which has been assumed to be the case in
Eq.\ (\ref{eqn:ks_eqns}).

The plane-wave DFT code \textsc{castep} \cite{castep} was modified to
allow the solution of Eq.\ (\ref{eqn:ks_eqns}).  We have used the LDA
exchange-correlation functional in all our calculations. It was
verified that the results were essentially unchanged when a
generalized-gradient-approximation functional was used instead.  To
minimize finite-size effects, the density at the edge of the cell
should be close to the target density, but the screening of the
positron implies that an additional electron is attracted to the
center of the cell.  To reduce the finite size effects we therefore
define the $r_s$ parameter via $(4/3)\pi r_s^3 = V/(N-1)$, where $V$
is the volume of the cell and $N$ is the number of electrons.  Our
calculations were performed in simple cubic cells and we used the
Ewald interaction to describe the Coulomb interactions.  We have also
restricted our attention to paramagnetic HEGs, although it is
straightforward to apply our method to spin-polarized HEGs.

\begin{figure}
\begin{center}
\includegraphics[clip,scale=0.4]{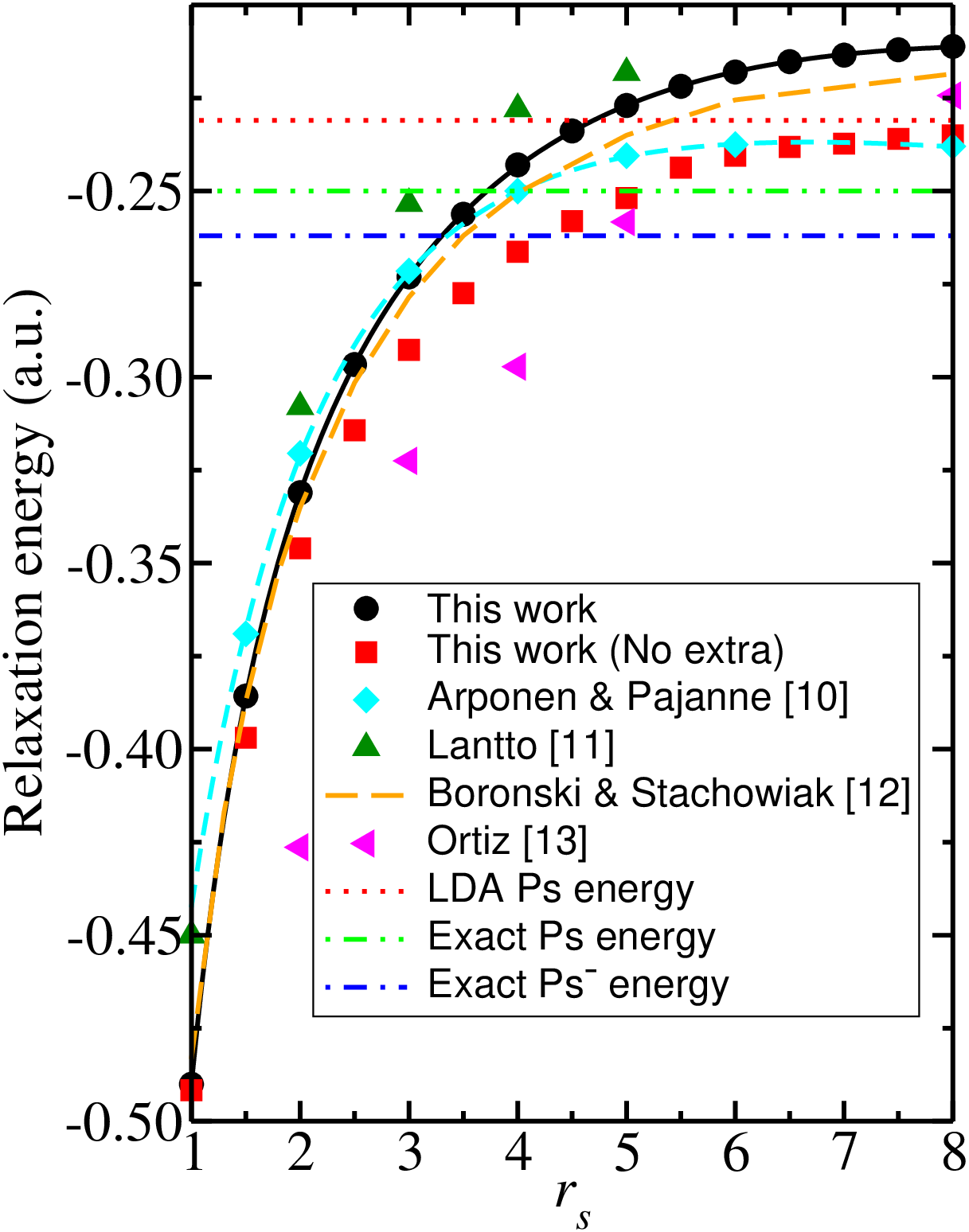}
\caption{(Color online) Relaxation energy against density parameter.
Results obtained with $N=114$, 162, and 246 electrons for $1 \leq r_s
\leq 3.5$, with $N=54$, 114, and 162 electrons for $4 \leq r_s \leq
6.5$, and with $N=14$, 54, and 114 electrons for $7 \leq r_s \leq 8$
were averaged to obtain the final results.  Also shown are the
relaxation energies obtained by other authors
\cite{arponen_1979,lantto_1987,boronski_1998,ortiz_1992}. The dashed
line following the data of Arponen and Pajanne \cite{arponen_1979} is
the widely used fit of Boro\'nski and Nieminen \cite{boronski_1986}.
The energy of the Ps$^-$ ion is taken from Ref.\ \cite{frolov_2005}.
\label{fig:final_relax}}
\end{center}
\end{figure}

Our relaxation energies were extrapolated to basis-set completeness
using the empirically derived expression $\Delta \Omega_{E_{\rm
    cut}}=\Delta \Omega_\infty +\kappa E_{\rm cut}^{-4/3}$, where
$\kappa$ is a fitting parameter and $E_{\rm cut}$ is the plane-wave
cutoff energy.  We estimate the residual finite-size errors in our DFT
relaxation energies to occur in the third significant figure.  The fit
to our data shown in Fig.\ \ref{fig:final_relax} is
\begin{equation}  
  \Delta \Omega(r_s) = \frac{A_{-1} r_s^{-1} +
    A_0+A_1r_s-0.262005B_2r_s^2}{1+B_1r_s+B_2r_s^2},
\end{equation} 
where $A_{-1}=-0.28877$, $A_0=-0.22339$, $A_1=0.011536$,
$B_1=0.012331$, and $B_2=0.020016$. This fitting form tends to the
energy of the Ps$^-$ ion at low density \cite{frolov_2005} and could
be used as the LDA electron-positron correlation functional in a
two-component DFT calculation for a positron in a real system.
Equation (\ref{fig:final_relax}) does not yield the exact high-density
behavior calculated within the random phase approximation, although
this is only relevant for $r_s<0.1$ \cite{arponen_1978}.

Our relaxation-energy results are in reasonable agreement with the
many-body-theory results of Refs.\ \cite{arponen_1979} and
\cite{boronski_1998}, but are in clear disagreement with the quantum
Monte Carlo (QMC) data of Ref.\ \cite{ortiz_1992}.  The orbitals in
the trial wave functions used in the QMC calculations reported in
Refs.\ \cite{ortiz_1992} and \cite{fraser_1995} were single plane
waves for the electrons and the positron, which do not allow for the
strong pairing that occurs between the electrons and the positron at
low density.  In common with the most accurate many-body-theory
results, our relaxation energy is higher than the energy of a Ps atom
at the lowest density considered of $r_s=8$.
Neglecting the extra interaction $\hat{V}_{\rm e}$ as suggested by
Leung \textit{et al.}\ \cite{leung} leads to relaxation energies that
increase monotonically towards the DFT-LDA energy of a Ps atom.  

The annihilating-pair MD depends sensitively on the accuracy of the
correlated electron-positron pairing $\phi_i({\bf x})$ and we expect
that our fully self-consistent treatment of the pairing orbitals will be
more accurate than previous approaches
\cite{kahana_1963,stachowiak_1990}.  The MD at each momentum $p$ was
extrapolated to basis-set completeness using the empirically
determined expression $\rho_{E_{\rm
    cut}}(p)=\rho_\infty(p)+\alpha(p)/E_{\rm cut}+\beta(p)/E_{\rm
  cut}^{3/2}$, where $\alpha(p)$ and $\beta(p)$ are fitting
parameters.  As the system size is increased, the momenta at which the
MD is defined become more finely spaced and the finite-size errors at
each point are reduced.  We have therefore fitted a model curve to our
MD data obtained at the largest system size available at each density.
We have verified that our results are well-converged with respect to
system size.  Below the Fermi wave vector $k_F$ we fit
$\rho(p)=w_0+w_2 p^2+w_4 p^4$ to our MD data, where $w_0$, $w_2$, and
$w_4$ are fitting parameters. Above $k_F$ the DFT-calculated MD falls
off exponentially and so we fit $\rho(p)=W \exp(-sp)$ to our data,
where $W$ and $s$ are fitting parameters.

\begin{figure}
\begin{center}
\includegraphics[clip,scale=0.4]{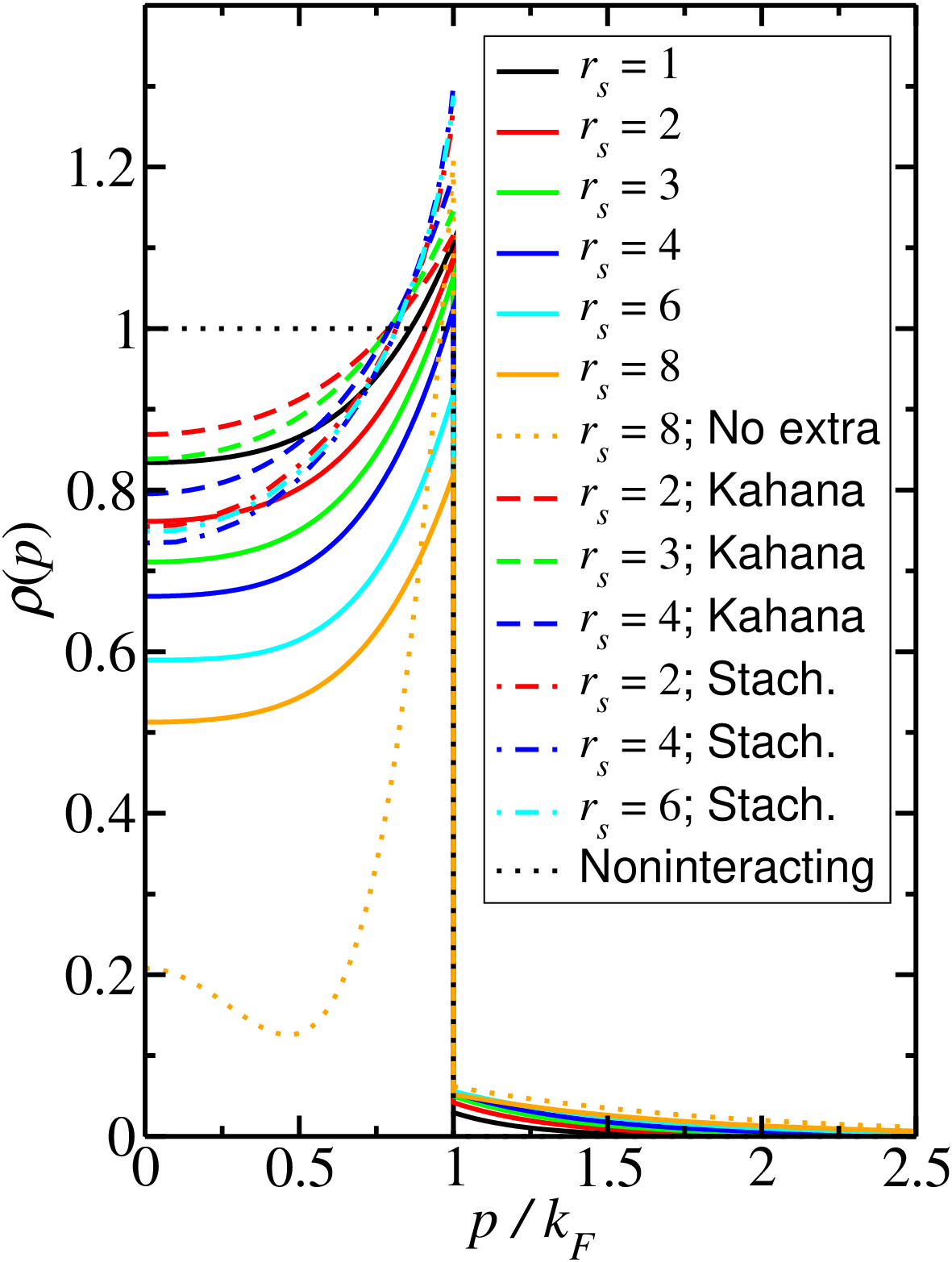}
\caption{(Color online) Annihilating-pair MDs $\rho(p)$ for different
  densities. $k_F$ is the Fermi wave vector.  The solid lines show the
  MDs calculated with the extra interaction included; the dotted line
  shows the MD calculated at $r_s= 8$ without the extra interaction.  
  Model functions have been fitted to the results obtained with $N=406$ 
  electrons at $r_s=1$, $N=246$ electrons for $2 \leq r_s \leq 7$, and 
  $N=162$ electrons at $r_s=8$.  MD data calculated by Kahana \cite{kahana_1963} 
  and Stachowiak \cite{stachowiak_1990} using other approaches are also shown.
\label{fig:final_md}}
\end{center}
\end{figure}

Electron-positron annihilating-pair MDs at different densities are
plotted in Fig.\ \ref{fig:final_md}.  The normalization is chosen such
that $\int_0^\infty 4 \pi p^2 \rho(p)\, dp = (4/3) \pi k_F^3$.  Our
results clearly show the enhancement of the annihilating-pair MD at
the Fermi edge predicted by Kahana \cite{kahana_1963}, but our MD data
differ quantitatively from the previous results
\cite{kahana_1963,stachowiak_1990} for $1 \leq r_s \leq 8$.  In this
range, we find the greatest enhancement of the MD at the Fermi edge at
$r_s=1$, whereas the previous works \cite{kahana_1963,stachowiak_1990}
found the enhancement of the MD to increase when the density is
lowered.  The previous works \cite{kahana_1963,stachowiak_1990} did
not report the weight in the MDs above the Fermi edge.  In our
calculations the exponential tail of the MD above the Fermi edge
carries an increasing amount of weight as the density is lowered,
which is responsible for the decrease in the enhancement that we find
at the Fermi edge.  We find the exponent $s$ to be of the order of
$r_s$ over the range of densities we have studied.  It is essential to
include the extra interaction when calculating the MD: omitting it
results in a dramatic increase in the enhancement at the Fermi edge
and substantial transfer of weight beyond the Fermi edge, as shown in
Fig.\ \ref{fig:final_md}.  The Kohn-Sham orbitals do not describe the
electron-electron correlation.  Such correlation effects (i) tend to
oppose the enhancement at the Fermi edge, particularly at low
densities, and (ii) introduce an algebraically decaying tail in the
MD\@.

The electron-positron PCF $g(r)$ is proportional to the charge density
in the frame in which the positron is stationary.  However, the fact
that we used a plane-wave basis set for our orbitals results in a
significant finite-basis error near $r=0$, because the Kimball cusp
condition \cite{kimball} is not satisfied.  We therefore extrapolated
the contact PCF to basis-set completeness by fitting our data to the
empirically determined expression $g_{E_{\rm
    cut}}(0)=g_\infty(0)+a/\sqrt{E_{\rm cut}}+b/E_{\rm cut}$, where
$a$ and $b$ are fitting parameters.  The residual finite-size error in
$g(0)$ is less than about 1\%.

\begin{figure}
\begin{center}
\includegraphics[clip,scale=0.4]{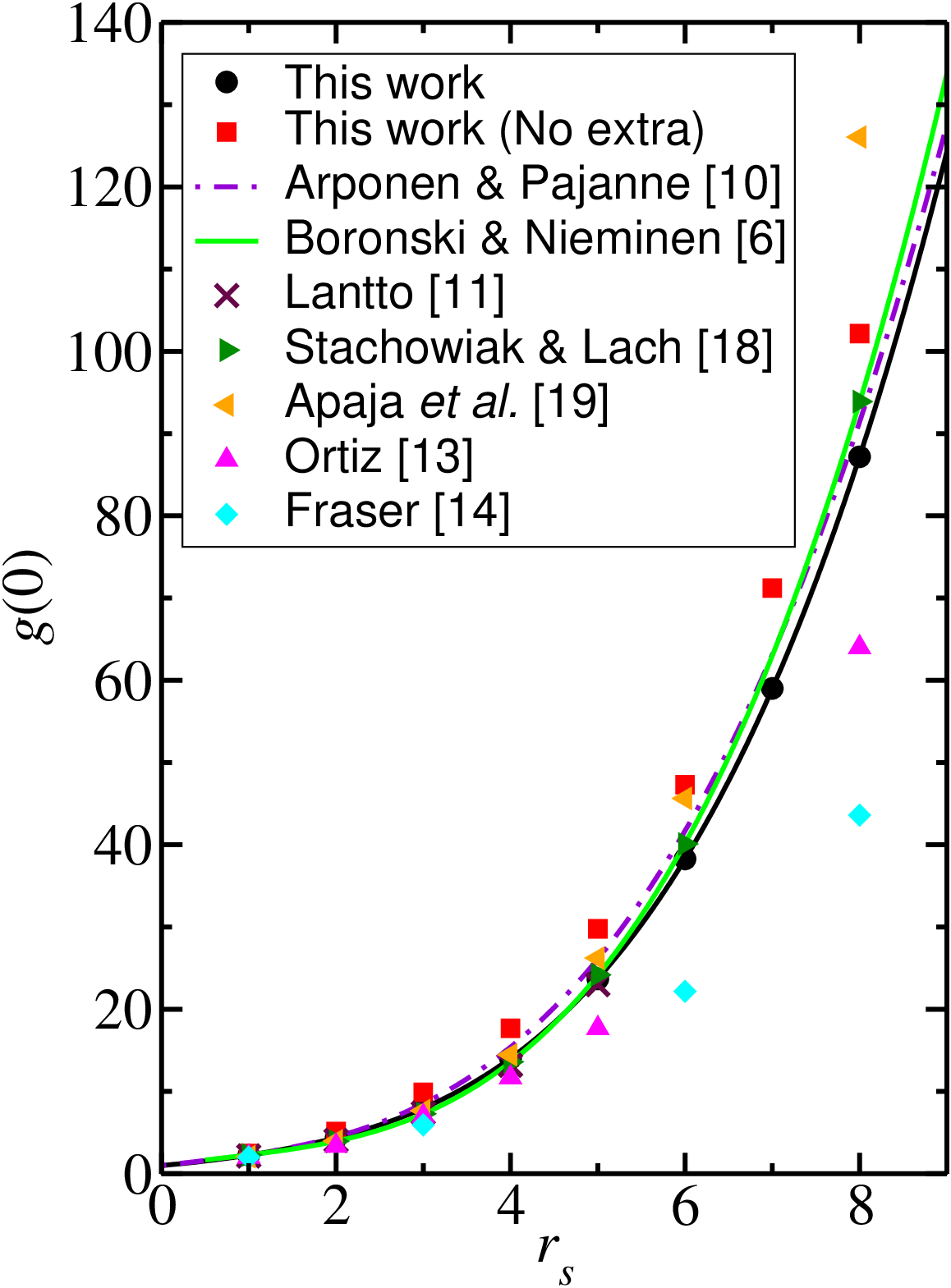}
\caption{(Color online) Electron-positron contact PCF $g(0)$ against
density parameter $r_s$.  The results were obtained by averaging the
$g(0)$ values obtained in cells with $N=114$, 162, and 246
electrons for $1 \leq r_s \leq 7$ and $N=114$ and 162 at $r_s=8$.
Various contact PCF results from the literature are also shown.
\label{fig:final_g0_v_rs}}
\end{center}
\end{figure}

The electron-positron contact PCF is plotted in Fig.\
\ref{fig:final_g0_v_rs}.  Adapting the fitting form of Ref.\
\cite{boronski_1986} slightly, we represent our contact PCF data by
\begin{eqnarray}  
  g(0) & = & 1+1.23r_s+a_{3/2}r_s^{3/2}+a_2r_s^2 +
  a_{7/3}r_s^{7/3} \nonumber \\ & & {} +a_{8/3}r_s^{8/3}+0.173694r_s^3,
\label{eqn:pcf_fit}
\end{eqnarray} 
where $a_{3/2}=-1.56672$, $a_2=4.16983$, $a_{7/3}=-3.579$, and
$a_{8/3}=0.836389$.  Equation (\ref{eqn:pcf_fit}) satisfies both the
high-density (random phase approximation) \cite{arponen_1978} and
low-density (Ps$^-$) limiting behavior \cite{frolov_2005}.

The extra interaction reduces the contact PCF at all the densities we
have studied, by about 8\% at $r_s=1$, rising to about 15\% at
$r_s=8$.  Our electron-positron PCF results are in reasonably good
agreement with the many-body-theory results of
\cite{arponen_1979,lantto_1987,stachowiak_1993,apaja_2003}, but are in
strong disagreement with the QMC results
\cite{ortiz_1992,fraser_1995}, which give much smaller values for the
reasons discussed earlier.

In summary we have used a modified one-component DFT code to calculate
the relaxation energy, contact PCF, and annihilating-pair MD of a
single positron in a HEG\@ by working in the frame in which the
positron is stationary.  It is interesting to observe that the data
required to parameterize a two-component exchange-correlation density
functional can be obtained using one-component DFT calculations.  Our
results for the positron relaxation energies and annihilation rates
are in broad agreement with earlier theoretical work based on
many-body-theory results.  The annihilating-pair MDs are particularly
sensitive to the description of the electron-positron correlation.  We
have reported MDs for a wider range of densities than previous
studies, and our calculations extend above the Fermi edge where we
find an exponential decay of the MD with momentum.

\begin{acknowledgments} 
  We acknowledge financial support from the Leverhulme Trust, Jesus
  College, Cambridge, and the UK Engineering and Physical Sciences
  Research Council (EPSRC)\@.  We thank P.J.\ Hasnip for assistance
  with the \textsc{castep} code.
\end{acknowledgments}


\begin{thebibliography}{99}

\bibitem{hohenberg_1964} P.\ Hohenberg and W.\ Kohn, Phys.\ Rev.\
\textbf{136}, 864 (1964).

\bibitem{kohn_1965} W.\ Kohn and L.J.\ Sham, Phys.\ Rev.\
  \textbf{140}, 1133 (1965).

\bibitem{gross_1961} E.P.\ Gross, Nuovo Cimento \textbf{20}, 454
(1961).

\bibitem{pitaevskii_1961} L.P.\ Pitaevskii, Sov.\ Phys.\ JETP
\textbf{13}, 451 (1961).

\bibitem{krause_rehberg} R.\ Krause-Rehberg and H.S.\ Leipner,
\textit{Positron Annihilation in Semiconductors}, Springer-Verlag
(1999).

\bibitem{boronski_1986} E.\ Boro\'nski and R.M.\ Nieminen, Phys.\
Rev.\ B \textbf{34}, 3820 (1986).

\bibitem{puska_1994} M.J.\ Puska and R.M.\ Nieminen, Rev.\ Mod.\
  Phys.\ \textbf{66}, 841 (1994).

\bibitem{rs_and_au} We use Hartree atomic units ($\hbar=|e|=m_{\rm
e}=4\pi\epsilon_0=1$) throughout.  Electron densities are specified by
the radius $r_s$ of the sphere that contains one electron on average
in units of the Bohr radius.

\bibitem{leung} C.H.\ Leung, M.J.\ Stott, and C.O.\ Almbladh, Phys.\
Lett.\ \textbf{57A}, 26 (1976).

\bibitem{castep} S.J.\ Clark, M.D.\ Segall, C.J.\ Pickard, P.J.\
Hasnip, M.J.\ Probert, K.\ Refson, and M.C.\ Payne, Z.\ f\"ur
Krystallographie \textbf{220}, 567 (2005).

\bibitem{frolov_2005} A.M.\ Frolov, Phys.\ Lett.\ A \textbf{342}, 430
(2005).

\bibitem{arponen_1978} J.\ Arponen, J.\ Phys.\ C \textbf{11}, L739
(1978).

\bibitem{arponen_1979} J.\ Arponen and E.\ Pajanne, Ann.\ Phys.\
\textbf{121}, 343 (1979).

\bibitem{lantto_1987} L.J.\ Lantto, Phys.\ Rev.\ B \textbf{36}, 5160
(1987).

\bibitem{boronski_1998} E.\ Boro\'nski and H.\ Stachowiak, Phys.\
Rev.\ B \textbf{57}, 6215 (1998).

\bibitem{ortiz_1992} G.\ Ortiz, PhD thesis, Swiss Federal Institute of
Technology, Lausanne (1992).

\bibitem{fraser_1995} L.\ Fraser, PhD thesis, Imperial College, London
(1995).

\bibitem{kahana_1963} S.\ Kahana, Phys.\ Rev.\ \textbf{129}, 1622
(1963).

\bibitem{stachowiak_1990} H.\ Stachowiak, Phys.\ Rev.\ B \textbf{41},
12522 (1990).

\bibitem{kimball} J.C.\ Kimball, Phys.\ Rev.\ A \textbf{7}, 1648
(1973).

\bibitem{stachowiak_1993} H.\ Stachowiak and J.\ Lach, Phys.\ Rev.\ B
\textbf{48}, 9828 (1993).

\bibitem{apaja_2003} V.\ Apaja, S.\ Denk, and E.\ Krotscheck, Phys.\
Rev.\ B \textbf{68}, 195118 (2003).

\end{thebibliography}
\end{document}